\documentclass[a4paper,oneside]{article}

\usepackage{microtype}
\usepackage[sc]{mathpazo}
\usepackage[T1]{fontenc}
\usepackage[DIV=11,BCOR=0cm,headinclude=true]{typearea}

\usepackage{amsfonts}
\usepackage{amsmath}
\usepackage{amssymb}
\usepackage{amsthm}
\usepackage{graphicx}
\usepackage{float}
\usepackage[config,labelfont=sc,textfont=sl]{caption,subfig}
\usepackage[british]{babel}
\usepackage[ruled,vlined,linesnumbered]{algorithm2e}

\usepackage[maxbibnames=100,autocite=superscript,style=custom-nature,sorting=none,defernumbers=true]{biblatex}

\usepackage{authblk}

\DeclareMathOperator*{\real}{Re}
\DeclareMathOperator*{\imag}{Im}
\DeclareMathOperator*{\sqroot}{sqrt}

\author[1,2]{Parry Y.\ Chen}
\author[2]{Yonatan Sivan}
\affil[1]{School of Physics and Astronomy, Raymond and Beverly Sackler Faculty of Exact Sciences, Tel Aviv University, Israel}
\affil[2]{Unit of Electro-optic Engineering, Ben-Gurion University, Israel}

\title{Robust Location of Optical Fiber Modes via the Argument Principle Method}
\date{\today}

\begin{document}
\maketitle

\begin{abstract}
We implement a robust, globally convergent root search method for transcendental equations guaranteed to locate all complex roots within a specified search domain, based on Cauchy's residue theorem. Although several implementations of the argument principle already exist, ours has several advantages: it allows singularities within the search domain and branch points are not fatal to the method. Furthermore, our implementation is simple and is written in MATLAB, fulfilling the need for an easily integrated implementation which can be readily modified to accommodate the many variations of the argument principle method, each of which is suited to a different application. We apply the method to the step index fiber dispersion relation, which has become topical due to the recent proliferation of high index contrast fibers. We also find modes with permittivity as the eigenvalue, catering to recent numerical methods that expand the radiation of sources by eigenmodes.
\end{abstract}

\section{Introduction}
A variety of numerical and analytical methods culminate in the solution of a transcendental equation,
\begin{equation}
f(z) = 0.
\label{eq:transcend}
\end{equation}
In electromagnetism and optics, dispersion relations are often defined as transcendental equations whose solutions lead to the eigenmodes of a structure. While conceptually simple, the required root search entails many implementational issues that damage the robustness of subsequent methods that rely on these eigenmodes, especially if complex roots are sought. Iterative methods such as Newton's method,
\begin{equation}
z_{n+1} = z_n - \frac{f(z_n)}{f'(z_n)},
\label{eq:newton}
\end{equation}
or Muller's method are frequently employed, based on an initial guess of the solution, $z_0$. These typically have excellent local convergence properties when initiated close to a root.\autocite{press2007numerical} However, their global convergence properties depend critically on the behavior of the attraction basins of the roots. While iterative methods will usually converge on to a solution, they may not converge to the desired solution and have difficulty reliably locating a complete set of solutions within the search domain.\autocite{press2007numerical,boyd2014solving}

\begin{figure}[tb]
\begin{center}
\includegraphics{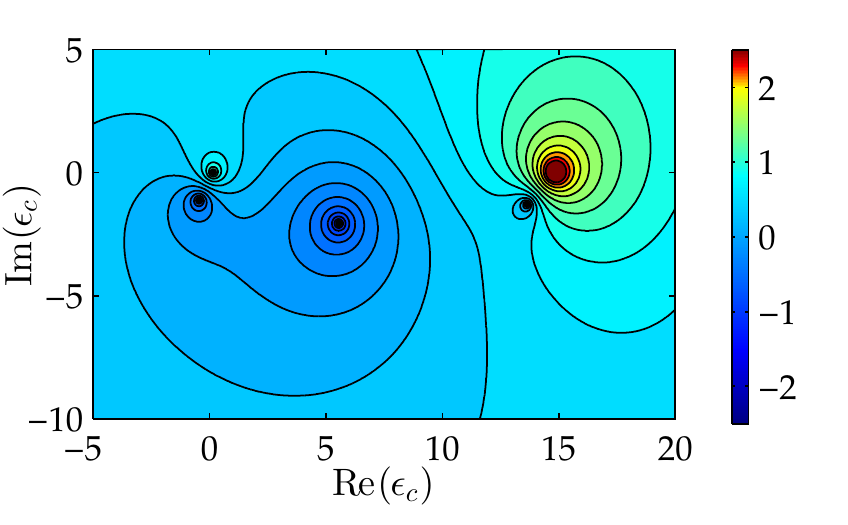}
\caption[]{Example attraction basins of three roots (deep blue) and two singularities (red). Horizontal and vertical axes represent real and imaginary parts of the complex search variable domain, while color is the logarithmic magnitude of the transcendental equation, $\log_{10}(|f(z)|)$. For methods such as Newton's method, each iteration moves in a direction roughly perpendicular to the level curves. Thus, the leftmost root exists within the attraction basin of the more dominant central root, while the rightmost exists in the shadow of the singularity. When Newton's method is initiated near these two roots, it either converges to the more dominant root or diverges from the singularity. The narrow attraction basins of these roots compound the difficulty, which can only be located with highly accurate initial guesses.}
\label{fig:basins}
\end{center}
\end{figure}

A number of features of optical dispersion relations can give rise to unfavorable attraction basins, impeding the success of iterative methods. Firstly, sharp roots with very narrow linewidths may be present, which often represent the more physically interesting modes since they have higher quality factors. These cannot be found unless an accurate initial guess is available. Secondly, roots can cluster, so that the attraction basin of one or more roots lies within the attraction basin of a more dominant root. Finally, dispersion relations feature both singularities and roots, and are commonly collocated. Since Newton's method diverges away from singularities, this masks the attraction basins of nearby roots. Example of these phenomena are plotted in Figure \ref{fig:basins}. For illustration purposes, we have deliberately chosen parameters with relatively tame attraction basins, since a more extreme example would have attraction basins too small to be visible in the figure. Such root search issues are by no means specific to optical dispersion relations, and can complicate iterative solutions to any complex transcendental equation \eqref{eq:transcend}. 

Variations of Newton's method are available, such as Davidenko's method, which are more tolerant of distant initial guesses.\autocite{davidenko1953new,hejase1993use} A related class of methods is the homotopy continuation methods, which exploits the property that a system of algebraic equations can be continuously deformed into a simpler system.\autocite{jimenez-islas2013nonlinear} However, much numerical experimentation is necessary to select the appropriate method and parameters for each system, and implementation complexity quickly escalates. Most importantly, none of the aforementioned methods guarantee that all roots of a general transcendental equation within a defined domain have been located, which is problematic if a complete set of modes is necessary.

Difficult attraction basins are nevertheless manageable if the roots sought are real. Robust bracketing techniques are available, which employ slow but dependable methods such as the bisection method in conjunction with Newton's method.\autocite{press2007numerical} Alternatively, \eqref{eq:transcend} can be expanded using Chebyshev polynomials, which does guarantee that all roots of the expansion will be found. A growing literature now exists on how to handle special points, such as singularities.\autocite{boyd2014solving} For roots in the complex plane however, few reliable methods exist. In the context of dispersion relations, complex eigensolutions emerge whenever the structures are lossy, or the eigenmodes sought are not bound to the structure. Indeed, it was our unsatisfactory experience with unreliable iterative methods which motivated our desire for a better alternative. Fortunately, the argument principle method provides a robust globally-convergent method capable of locating all roots within a specified region of the complex plane, without needing to supply initial guesses and regardless of the behavior of attraction basins.\autocite{delves1967numerical,boyd2014solving} Its robustness is due in part to Cauchy's theorem, which provides a count of the number of roots enclosed by a given contour. It is widely applicable, and requires only that the specified search domain be meromorphic, that is, free of branch cuts and essential singularities.


Several freely available implementations of the argument principle method exist,\autocite{botten1983complex,kravanja2000zeal} as well as related methods,\autocite{gillan2006computing} which are applicable to a wide class of transcendental equations. However, many variations of the argument principle method are possible, discussed in greater detail in Section \ref{sec:method}, and specific choices must be made for each numerical implementation. Two key aspects of these preexisting implementions disqualify them for use with our target dispersion relation. Firstly, they assume that the desired search domain is analytic, and so is free of even isolated singularities. Secondly, these methods subdivide and enlarge the search domain to overcome numerical issues encountered by the argument principle method. While this is unproblematic for analytic domains, our dispersion relation features singular branch points, which induce these implementations to enlarge their contours to include these non-meromorphic points and consequently fail. These choices, especially the subdivisioning methodology, are integral to the implementations, and cannot be easily modified.

Additionally, the existing implementations are all written in FORTRAN.\autocite{botten1983complex,kravanja2000zeal,gillan2006computing} However, the speed advantage of FORTRAN is inconsequential to the argument principle method, as it is not at all numerically demanding relative to modern computing resources. Of greater importance is customizability, given the multitude of innovations and variations that have been developed. The dated FORTRAN 77 style of the existing implementations impedes this process. This underscores the need for a freely available and easily extensible implementation written in a rapid development programming language, ready to be modified and integrated to suit the needs of each individual application. We have chosen MATLAB for this purpose, due to its prevalence in scientific computing.


Numerous specializations of the argument principle method have been developed for optical dispersion relations, mostly applicable to the modes of multi-layer planar waveguides.\autocite{smith1991numerical,anemogiannis1992multilayer,smith1992mode,kwon2004simple,gillan2006computing,kocabas2009modal} Few have applied the method to the dispersion relation for circular step-index fibers. In the past, solving the full dispersion relation was largely unnecessary, avoiding the associated complications.\autocite{snyder1983optical,okamoto2006fundamentals}
Instead, the fiber modes were obtained from either a simpler approximate dispersion relation or various asymptotic expansions, enabled by the dominance of single mode fibers with weak index contrast in telecommunications optics. However, use of low index contrast fibers was largely born of necessity, due to the difficulty of drawing fibers with thermally incompatible components, so core and cladding could only be differentiated by chemical doping.\autocite{ballato2013rethinking,tao2015multimaterial}

Recently, significant research attention has been devoted to drawing fibers composed of different glasses, or combinations of glass, plastic, semiconductor, and metal. A wealth of new applications are possible, offering not only better material properties for optical propagation and optical nonlinearities, but also incorporating piezoelectric, electric, electronic, and optoelectronic functionalities onto the fiber. For recent reviews, see Refs \parencite{ballato2013rethinking,tao2015multimaterial,schmidt2016hybrid}. For example, chalcogenide glasses can be encased in a polymer cladding to exploit their strong nonlinearities,\autocite{eggleton2011chalcogenide} while mitigating their weak mechanical strength.\autocite{tao2012multimateriala} The strong index contrast helps confine the mode to the chalcogenide core, further enhancing nonlinearities. Silicon cores are desirable for their high damage threshold, heat dissipation, and transparency across the infrared,\autocite{ballato2008silicon} while metallic fibers provide electrical conductivity, nanoscale optical confinement, and large polarization discrimination.\autocite{spittel2015curvature,jain2016silver,tuniz2016broadband} The preceding examples feature a single interface, but the robust nature of the argument principle method is also ideal for fiber dispersion relations involving multiple concentric interfaces. This is particularly applicable to hollow core Bragg fibers,\autocite{yeh1978theory} useful for their ability to tune field profiles, deliver high intensity beams, and perform sensing on nanovolume samples.


In addition to understanding the source-free propagation along fibers, another motivation for solving the full fiber dispersion relation comes from modeling the emission of electrodynamic sources. In other words, solutions to the inhomogeneous Maxwell equations are sought, rather than the above homogeneous examples. Several recent methods expand fields radiated by sources in terms of the source-free eigenmodes of the nearby cylindrical structure.\autocite{bergman1980theory,sauvan2013theory,kristensen2014modes,farhi2016electromagnetic} This approach holds many advantages, providing closed form expressions for enhancement of spontaneous emission by lossy and leaky resonators, and allowing rapid extraction of the spectral and spatial dependence of emitters with the one set of eigenmodes. Eigenmodes with either permittivity or frequency as their eigenvalue can be employed, with the latter known as quasi-normal-modes. In both cases, complex eigenvalues are usually required, even for lossless structures, since the structures simulated are leaky. Moreover, completeness of the eigenmodes is vital to ensuring correct expansion of fields. These methods are applicable to both dielectric and metallic cylindrical resonators, which typically have large and small radii respectively.\autocite{righini2011whispering,chang2007strong} The cylinder modes can couple strongly to emitting molecules and other optical sources, with applications to quantum optics and sensitive sensing. Finally, these eigenmode methods can extend applicability of the fiber dispersion relation, valid for a single core-cladding interface, to find eigenmodes of photonic crystal fibers without further simulation.

We implement the argument principle method for the step index fiber dispersion relation. Our method can return eigenmodes corresponding to either complex propagation constant or complex permittivity, suitable for describing source free propagation and expanding fields of sources. Both bound and leaky modes can be found. The latter are modes that diverge at infinity and satisfy yet another source of need: they are convenient substitutes for the continuum of radiation modes,\autocite{hu2009understanding} and provide a basis for lasing theory beyond closed cavities.\autocite{tureci2006self} In some contexts, the radiative modes are also given by the dispersion relation,\autocite{pfeiffer1974surfacea} which can also be found using our method. With minor changes, our method can return complex frequencies, or any other variable as the eigenvalue. Since many variations of the argument principle method exist, our selections are motivated by simplicity and robustness of our resulting implementation. In particular, the method should reliably return all solutions of the dispersion relation, while avoiding failure due to interaction with branch points. Speed is a secondary consideration, but is less important due to the small cost of evaluating the dispersion relation. High impact speed optimizations are implemented, especially during numerical quadrature.

It should be emphasized though that our argument principle implementation is general to any transcendental equation \eqref{eq:transcend}. It is particularly suited to problems which have singularities in the search domain, whose contributions can be deflated if their locations are known. It is also suited to problems where the contours used can or should be predetermined, for example, to avoid problematic areas of the search domain. We implement the standard argument principle method, which requires the derivative of \eqref{eq:transcend}. Since this can be burdensome or impossible to obtain, we also implement a simple derivative free contour method, suitable for equations where all roots are known to be simple.

The paper is organized as follows. Section \ref{sec:disprel} introduces the dispersion relation for the step index fiber, and discusses a few of its general characteristics. Section \ref{sec:method} introduces the established literature on the argument principle method, detailing the many variations of the method and evaluating the suitability of each for the fiber dispersion relation. Section \ref{sec:implement} presents the details of our numerical implementation, discussing how the contours are chosen, and how the singularities and branch points are handled. Section \ref{sec:numerics} presents the dispersion relations obtained by the method, for both complex propagation constant and complex permittivity eigenvalues. Finally, Section \ref{sec:conclusion} concludes.

\section{Dispersion relation and modes}
\label{sec:disprel}
The step-index fiber dispersion relation defines a self-sustaining mode of a cylinder that exists in the absence of incident energy,\autocite{snyder1983optical,okamoto2006fundamentals,pfeiffer1974surfacea}
\begin{equation}
\left(\frac{\mu_c}{\alpha_c a}\frac{J'_m(\alpha_c a)}{J_m(\alpha_c a)} - \frac{\mu_b}{\alpha_b a}\frac{H'_m(\alpha_b a)}{H_m(\alpha_b a)}\right) \left(\frac{\epsilon_c}{\alpha_c a}\frac{J'_m(\alpha_c a)}{J_m(\alpha_c a)} - \frac{\epsilon_b}{\alpha_b a}\frac{H'_m(\alpha_b a)}{H_m(\alpha_b a)}\right) - \left(\frac{m\beta}{k}\right)^2 \left(\frac{1}{(\alpha_c a)^2}-\frac{1}{(\alpha_b a)^2}\right)^2 = 0.
\label{eq:disprel}
\end{equation}
Here, $J_m$ denotes the Bessel function of the first kind, and $H_m \equiv H_m^{(1)}$ is the Hankel function of the first kind, unless otherwise specified. Also, $m$ defines the angular variation of the field $e^{im\phi}$, $a$ is the radius of the cylinder, $\epsilon$ and $\mu$ are the relative permittivities and permeablities of the cylinder $c$ and background $b$, $k$ is the wavenumber, and $\beta$ is the longitudinal propagation constant. The in-plane propagation constants differ between the cylinder and background,
\begin{align}
\alpha_c^2 &= k^2\epsilon_c\mu_c - \beta^2, & \alpha_b^2 &= k^2\epsilon_b\mu_b - \beta^2.
\label{eq:alpha}
\end{align}

Eigenvalues of the dispersion relation can be defined for any of the variables $k$, $\beta$, $\epsilon$, or $\mu$, maintaining the remaining variables fixed. However, in this document, we restrict attention to root searches in the variables $\beta$ and $\epsilon_c$. The former usually corresponds to bound modes propagating along the fiber, the latter to eigenmodes useful for expanding nearby sources.

We consider in this paper solutions to the full dispersion relation, \eqref{eq:disprel}. However, we remark that in practice a number of simplifications to \eqref{eq:disprel} may be employed instead to find approximate solutions. As mentioned, simplifications are possible when the index contrast is low.\autocite{snyder1983optical,okamoto2006fundamentals} Meanwhile, the Bessel functions in \eqref{eq:disprel} can be expanded in either the large argument or small argument limits, which relate to the geometric optics and long wavelength limits respectively. The latter for example gives accurate results for the fundamental mode when it asymptotes towards the light line.\autocite{chang2007strong} Thus, \eqref{eq:disprel} is necessary only for high index contrast fibers in the intermediate frequency regime, which nevertheless covers a wide range of applications.

\section{Argument principle method}
\label{sec:method}
A substantial body of literature now exists on the argument principle method in all its variations. In this section we outline the argument principle method, as first proposed by Delves and Lyness.\autocite{delves1967numerical} This then permits us to introduce the many developments since the original publication. We assess the suitability of each in the context of solving the the step-index fiber dispersion relation \eqref{eq:disprel}, and justify the choices made in our implementation.

The argument principle is based on Cauchy's theorem, and counts the number of roots $R$ and poles $P$ within a closed contour $\partial C$ in the complex plane,
\begin{equation}
\frac{1}{2\pi i} \oint_{\partial C} \frac{f'(z)}{f(z)}\, dz = R-P.
\label{eq:argprin}
\end{equation}
In \eqref{eq:argprin} and in all further equations, a count of the roots and poles includes their multiplicities. To then locate the positions of the roots, the integrand of \eqref{eq:argprin} is projected onto a basis of functions, which acts as coordinates for the roots. Following Delves-Lyness, we use the simplest possible basis, the powers of $z$, to define the moments
\begin{equation}
s_n = \frac{1}{2\pi i} \oint_{\partial C} z^n\frac{f'(z)}{f(z)}\, dz = \sum_{k=1}^R z_k^n - \sum_{l=1}^P z_l^n,
\label{eq:moments}
\end{equation}
where $z_k$ are the locations of the roots and $z_l$ are the locations of the poles. Again following Delves-Lyness, $R$ moments need to be evaluated to ensure there are sufficient linearly independent equations to obtain the location of each root from the set of moments, $\{s_n\}$. Numerical implementation of \eqref{eq:argprin} and \eqref{eq:moments} proceeds through use of an appropriate numerical quadrature rule.

Ordinarily, the negative contributions of poles to \eqref{eq:argprin} and \eqref{eq:moments} prevent further progress. Fortunately, the locations and orders of all poles of many transcendental equations are known analytically, including our target dispersion relation \eqref{eq:disprel}. This is true because singularities usually originate from individual terms while the roots require balance between all terms. Thus, the negative contributions of the known singularities can be deflated, leaving only the positive contributions of the roots. Previously found roots can also be deflated, to yield deflated moments
\begin{equation}
s_n^* = s_n - \sum_{k=1}^D z_k^n + \sum_{l=1}^P z_l^n = \sum_{k=1}^N z_k^n,
\label{eq:deflate}
\end{equation}
where $D$ is the number of roots to deflate and $N$ is the remaining roots. However, this ability to deflate singularities is missing from freely available implementations, which renders them inapplicable to \eqref{eq:disprel}.\autocite{botten1983complex,kravanja2000zeal,gillan2006computing}

The argument principle method relies on the domain $C$ being meromorphic, in other words, free of essential singularities and branch cuts. While the the fiber dispersion relation \eqref{eq:disprel} is free of essential singularities, branch points can appear due to the square root and Hankel functions. In fact, these branch points coincide and correspond to the light line, though the branch cuts that emanate may radiate in different directions. Several strategies for treating the square root branch point were discussed in a series of papers by Smith et al.\autocite{smith1991numerical,smith1992mode,smith1993unfolding} Another more mathematically sophisticated method is to perform a logarithmic conformal mapping of the complex plane, placing the branch point at infinity.\autocite{bakhtazad1997general} This not only unfolds the two Riemann sheets of the square root function, but can also unravel the infinite number of Riemann sheets of the Hankel function and unite them onto the one complex plane. However, logarithmic transformations distort the complex plane, and in particular compresses the otherwise regular spacing between the singularities. This complicates the selection and placement of contours. Since only a single branch point exists in \eqref{eq:disprel}, we implement a simpler strategy of adjusting the contours, to be discussed in Section \ref{sec:branch}.

Once the moments have been obtained and deflated, the Delves-Lyness method concludes by constructing a polynomial with the same complex roots as the original transcendental equation,
\begin{equation}
\prod_{k=1}^N (z-z_k).
\label{eq:poly}
\end{equation}
Newton's identities are used to relate the moments \eqref{eq:deflate} to \eqref{eq:poly}, which can be solved by well known methods for polynomials.\autocite{press2007numerical} Appendix \ref{sec:newid} describes this procedure in detail. However, the solution of high order polynomials is very numerically sensitive, so if too many roots are enclosed within the contour, the solution suffers from round-off errors and the resulting inaccuracy becomes unacceptable. Delves and Lyness suggested to mitigate this issue by enclosing no more than 5 roots per contour after deflation. A count of the number of enclosed roots is provided by \eqref{eq:argprin}. The search domain can be iteratively subdivided into smaller and smaller contours until the condition of 5 or fewer roots is met, allowing the polynomial \eqref{eq:poly} to be constructed and solved. Various subdivision shapes have been employed, such as overlapping circles, tessellating squares or annular sectors.\autocite{botten1983complex} More sophisticated methods use adaptive subdivisioning based on Delaunay triangulation, after first sampling the function value across the search domain.\autocite{kowalczyk2015complex} This carries the advantage of adapting the triangular mesh to avoid branch cuts. Finally, in some variations, subdivisioning continues until only a single root is enclosed, thus avoiding even the need to construct \eqref{eq:poly}.\autocite{dellnitz2002locating}

While the repartitioning of the search domain solves the problem of numerical sensitivity of high order polynomials, another issue arises. When contour boundaries are situated too close to a root or singularity, the evaluation of the moments \eqref{eq:moments} via numerical quadrature becomes slowly convergent. This occurs because the integrand of the argument principle \eqref{eq:argprin} resembles a simple pole near such points. The sharp peak that appears along the contour can be detected during numerical quadrature and, if encountered, a simple strategy is to abandon numerical quadrature on the current contour, and choose a slightly larger contour to evaluate. This procedure is adopted in several implementations,\autocite{delves1967numerical,kravanja2000zeal,gillan2006computing} but suffers from several drawbacks. Firstly, it can lead to failure if the contour passes through the vicinity of several roots.\autocite{gillan2006computing} However, the more pertinent disadvantage for our purposes is that the enlarged contour may cross a branch point or branch cut, which is fatal to the method. This is a common occurrence for \eqref{eq:disprel}, since eigenmodes often exist close to the light line, where a singular branch point exists. Thus, these numerical implementations are unsuitable to use, and another method of managing both the numerical sensitivity and slow convergence is necessary.

The aforementioned implementations which continually repartition the search domain are general ``black-box'' methods, suitable for problems where no further information on the behavior of the solutions is available. However, we are able to build a simpler and more robust implementation using prior knowledge of the transcendental equation to be solved, \eqref{eq:disprel}. Thus, can select our contours from the outset, and avoid the need to repartition and all its associated problems. Such contours each contain no more than 4 roots, yet the set of all contours will enclose all the roots but none of the branch cuts or branch points. Slow convergence during numerical quadrature due to contours lying close the branch point or roots is managed by using adaptive quadrature instead. This simplifies the implementation of the method, and yet renders it robust against branch points. The details are described in full in Section \ref{sec:implement}.

We remark that by incorporating knowledge of our dispersion relation into the contour selection process, we have specialized our method to this transcendental equation. However, the physical insight gained into the behavior of the roots is invaluable. Since only knowledge of the locations of singularities and branch points was required for contour selection, and this knowledge is already necessary in order to use the argument principle method, it was well worth the small additional effort of tailoring the contours to the transcendental equation. We emphasize though that once the contours have been selected, our underlying implementation of the argument principle method is general to any meromorphic transcendental equation.

Alternatively, several techniques have been developed in the literature to avoid altogether the numerical sensitivity of the polynomial \eqref{eq:poly}. Firstly, the process of retrieving the roots from the moments can be achieved by homotopy continuation methods, since the argument principle method \eqref{eq:moments} converts a general transcendental equation into a highly structured polynomial system of equations, which is more amenable to homotopy methods.\autocite{li1983locating} Secondly, the moments \eqref{eq:moments} can be evaluated using formal orthogonal polynomials as weight functions, subsequently retrieving the roots using a generalized linear eigenvalue problem to be described in Appendix \ref{sec:geneig}.\autocite{kravanja1999locating} These methods are also less sensitive to closely spaced and multiple roots. Finally, the need to evaluate many moments can be avoided by operating on the $2 \times 2$ matrix form of the dispersion relation \eqref{eq:disprel}, using instead matrix equivalents of the argument principle method.\autocite{sato2015extraction}  However, these methods add unnecessary complexity to the numerical implementation, as we achieve identical outcomes for our purposes by judicious selection of contours.

Another disadvantage of the original Delves-Lyness method is the need for the derivative $f'(z)$ in \eqref{eq:argprin} and \eqref{eq:moments}. Even though an explicit derivative is available for our target dispersion \eqref{eq:disprel}, we still consider and implement a derivative free method for future use with dispersion relations where explicit derivatives may be more difficult or tedious to obtain, for example, multilayered cylinders. Delves and Lyness' original proposal is to replace the integrand of the argument principle with $\log(f(z))$, but tracking the current sheet of the complex logarithm can pose difficulties.\autocite{delves1967numerical} For some practical purposes, a numerical estimate of $f'(z)$ based on $f(z)$ is adequate.\autocite{austin2014numerical} 

The most robust and accurate derivative-free method is due to Kravanaja and van Barel.\autocite{kravanja1999derivative,austin2014numerical} If all roots are known to be simple, which is the case for \eqref{eq:disprel} and many dispersion relations in general, a particularly simple form is available,
\begin{equation}
t_n = \frac{1}{2\pi i} \oint_C z^n\frac{1}{f(z)}\, dz = \sum_{k=1}^R \frac{z_k^n}{f'(z_k)}.
\label{eq:derivfree}
\end{equation}
The locations of the roots are now weighted by the function derivative, but the locations can still be retrieved by solving a generalized eigenvalue problem, detailed in Appendix \ref{sec:geneig}.

Besides not requiring the function derivative, \eqref{eq:derivfree} has several advantages over \eqref{eq:moments}. Firstly, the moments are insensitive to any singularities inside the domain $C$, so the location and deflation of singularities is not required. Furthermore, since the roots are now weighted by $f'(z)$, closely spaced roots are easily distinguished, avoiding numerical sensitivity issues. However, roots with larger derivatives now have less weight and are located with less accuracy. This is a disadvantage since it is precisely these roots that have narrower attraction basins, and require greater accuracy. Furthermore, deflation of known roots, while possible, is not as straightforward. Finally, no reliable derivative-free method exists for counting the number of roots enclosed in a contour, unlike \eqref{eq:argprin}. Thus, use of \eqref{eq:moments} is preferred if an explicit derivative is available.

Finally, roots found using the argument principle method are usually polished with an iterative method such as Newton's method. This hybrid approach combines the rapid local convergence of iterative methods with the robust global convergence of the argument principle method. We implement both Newton's method and the secant method, depending on whether or not the derivative is available. Alternatively, rational function interpolation is also possible, which may be more suitable for roots near singularities.

\section{Numerical implementation}
\label{sec:implement}
Our numerical implementation of the argument principle method consists of four parts: (a) identifying the locations and orders of singularities within the search domain, (b) selection of contours based on the singularities, (c) adjustment of the contours based on branch points, (d) discretization and numerical evaluation of the moments \eqref{eq:moments} or \eqref{eq:derivfree}, and (e) retrieval of the roots from the moments using Newton's identities \eqref{eq:newid} or the generalized eigenvalue equation \eqref{eq:geneig}. High level psuedocode summarizing the selection of contours and their use with the argument principle method is described in Algorithm \ref{alg:contour}, while the argument principle method itself is described in Algorithm \ref{alg:argprin}. First however, rationale on how the contours are chosen is discussed in Section \ref{sec:selection}, before continuing with the details of the numerical implementation in Sections \ref{sec:sings}--\ref{sec:discrete}.

\subsection{Contour considerations}
\label{sec:selection}
Several considerations exist in the choice of contours, all motivated by the desire to create a robust but simple implementation of the argument principle method for the step index fiber dispersion relation \eqref{eq:disprel}. To this end, we select our contours from the outset, informed by knowledge of the behavior of \eqref{eq:disprel}, rather than continually subdividing the search domain. Each contour should enclose 5 or less roots to avoid numerical sensitivity issues, and must not contain any branch points or cross any branch cuts. Since the argument principle method was selected for its robust global convergence, it is paramount that the set of all contours contains all the roots. Finally, it is desirable for rapid numerical convergence that the contour not be situated near any singularities.

Our selection of contours which satisfy all these criteria relies on a key assumption: for each singularity associated with $\alpha_c a$ in \eqref{eq:disprel}, counting multiplicity, there exists a root of the dispersion relation nearby. This is true for bound modes of lossless fibers in the weakly guiding limit, where a simplified dispersion relation applies.\autocite{snyder1983optical,okamoto2006fundamentals} Here, each singularity is a simple pole on the real axis, with neighboring poles each bracketing one root on the real axis. An analogous result holds for lossless bound modes of \eqref{eq:disprel}, where double poles now appear and bracket two roots. From repeated numerical experimentation, we observe that even as the roots become complex, they do not stray far from their neighboring singularities. Thus, the locations of the singularities, known analytically, provide guideposts for the locations of the roots. An infinite number of singularities exist, beginning at $\alpha_c a = 0$. Correspondingly, an infinite number of roots exist, with the fundamental radial mode of each angular order associated with the $\alpha_c a = 0$ singularity.

This knowledge permits a simple contour selection procedure which satisfies four of the five criteria prescribed above. Firstly, a set of contours centered on the set of singularities captures all of the roots. The number of contours to choose is dictated by the number of roots desired, and the completeness of the roots found can subsequently be confirmed by a simple count. Given that the majority of the singularities of \eqref{eq:disprel} are double poles, a contour that encloses only one unique singularity and excludes neighboring singularities should enclose at most 4 roots. Finally, by centering the singularities within each contour, they will not interfere with the numerical quadrature of the contours. Numerical details of the contour selection are provided in Section \ref{sec:sings}, where the locations of the singularities are also found.

The only outstanding consideration is the exclusion of branch points and branch cuts. Only one branch point exists, even though the dispersion relation \eqref{eq:disprel} has the potential for three. Two of the branch points coincide, and the third is a square root branch point which fails to manifest. We can avoid this solitary branch point by one simple adjustment to the contours. The two branch cuts that originate from the branch point can also be adjusted to avoid conflict with the contours. Finally, several singularities exist within close proximity of the branch point, and the branch point itself is singular. This causes slow convergence for contours in the search of roots that lie nearby, which can be resolved by switching to adaptive quadrature. Details are given in Section \ref{sec:branch}.

Finally, we discuss the shape of our contours. Since shape is not critical to the success of the method, we choose to use circular contours, which are the simplest to implement. Its main advantage is the that contours are periodic, so the equidistant trapezoidal rule converges exponentially. This efficiency compensates for the need for overlapping circles to cover the search domain. Finally, if the aspect ratio of a circle is inappropriate for covering the domain, it can be stretched or compressed into an ellipse, while still maintaining the periodicity of the contour.


\subsection{Singularities and their order}
\label{sec:sings}
Locating the poles of the dispersion relation \eqref{eq:disprel} and identifying their orders is an important task which allows their contributions to the moments to be deflated using \eqref{eq:deflate}, and determines the location of the contours. Four types of singularities exist, the two most important ones being those associated with the zeros of $\alpha_c a$ and $J_m(\alpha_c a)$ in the denominators of \eqref{eq:disprel}. The latter are simpler to analyze, and lead to double poles. To see this, we first ignore the special case of the zero of $J_m(z)$ at $z=0$. There exists an infinite number of zeros of $J_m(z)$, all simple and all on the positive real axis, denoted by $j_{m,l}$.\autocite{abramowitz1964handbook} The double appearance of $J_m(\alpha_c a)$ in \eqref{eq:disprel} then leads to double poles at
\begin{equation}
k^2 \epsilon_c\mu_c - \beta^2 = \left(\frac{j_{m,l}}{a}\right)^2.
\label{eq:sings}
\end{equation}

The singularity at $\alpha_c a = 0$ coincides with zeros of $J_m(\alpha_c a)$ and $J'_m(\alpha_c a)$ at $\alpha_c a=0$. For $m \neq 0$, this leads to a simple pole at $\alpha_c a = 0$, while for $m = 0$ no pole exists. This follows from small argument expansion for $J_m(z)$ and its derivative valid for $m > 0$,\autocite{abramowitz1964handbook}
\begin{align}
J_m(z) &\sim m! \left(\frac{z}{2}\right)^m, & J'_m (z) &\sim \frac{m}{z} J_m(z).
\end{align}
Thus, to leading order, the dispersion relation \eqref{eq:disprel} becomes
\begin{equation}
\frac{m\mu_c}{(\alpha_c a)^2} \frac{m\epsilon_c}{(\alpha_c a)^2} - \left(\frac{m\beta}{k}\right)^2 \frac{1}{(\alpha_c a)^4} = 0,
\end{equation}
so the singularities cancel. This unmasks the simple pole that exists in \eqref{eq:disprel} at $\alpha_c a = 0$. Meanwhile, for $m=0$, $J_0(z)$ does not have a root at $z=0$, and $J'_0(z) = -J_1(z)$ has a simple root at $z=0$ which cancels the root at $\alpha_c a = 0$. Thus, no pole exists. In passing, we note that this is the mathematical origin for the $m=1$ mode usually being the fundamental mode, rather than $m=0$.

Note that despite the square root in the definition of $\alpha_c$, no branch point exists at $\alpha_c a = 0$. The expression $\alpha_c a$ only appears in even functions in \eqref{eq:disprel}, so the sign of the square root ordinarily associated with its two Riemann sheets is inconsequential. For example, this is evident in the term $J'_m(z)/zJ_m(z)$, since $J_m(z)$ is always odd or even, depending on $m$, while $J'_m(z)$ always has the opposite parity, which ultimately cancels the odd parity of $z$.

The two remaining types of singularities, due to the roots of $\alpha_b a$ and $H_m(\alpha_b a)$, are less important to consider. When $\alpha_b a = 0$, a singularity is present, however this coincides with the branch point representing the light line. Discussion of this branch point and its impact on contour selection is deferred until Section \ref{sec:branch}. The final type of singularity originates from the complex zeros of $H_m(z)$. Fortunately, the majority of these are eliminated when the branch cut of $H_m(z)$ is relocated in Section \ref{sec:branch}, so treatment of the few remaining singularities is also deferred until then.

\begin{figure}[tb]
\begin{center}
\includegraphics{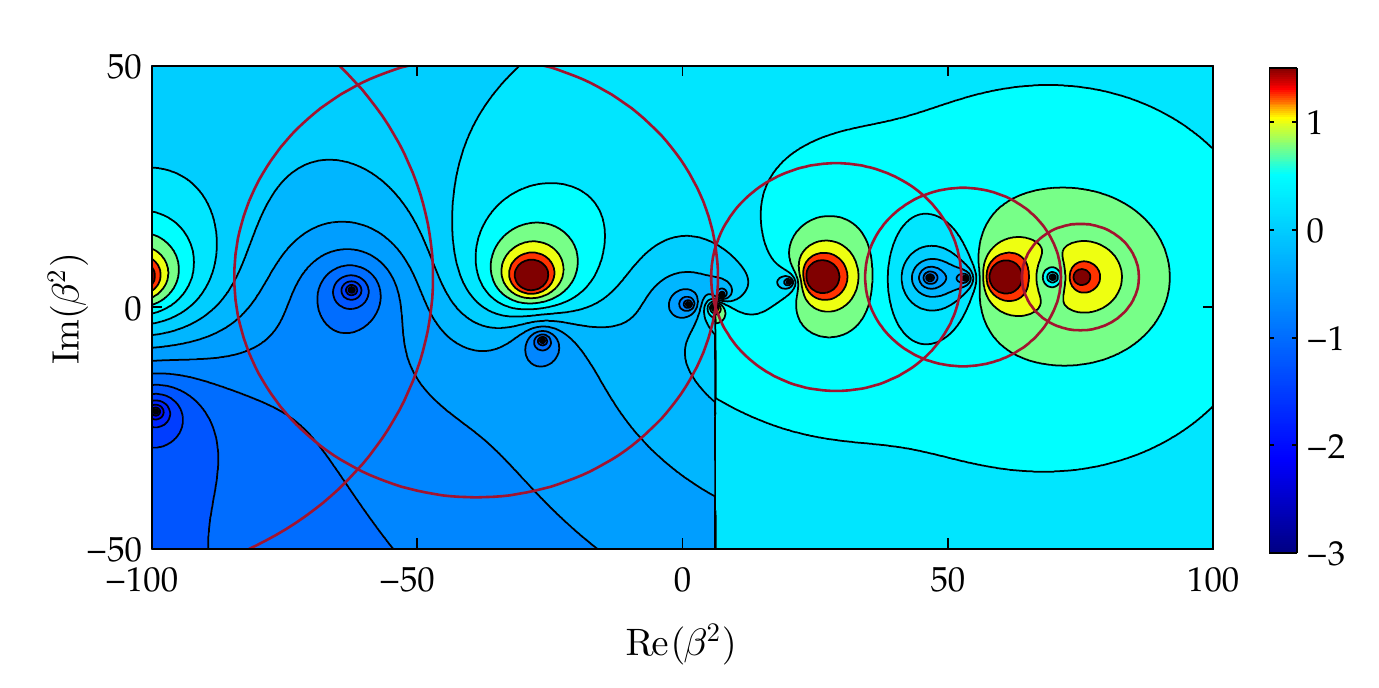}
\caption[]{Illustration of how contours are chosen and modified according to Sections \ref{sec:sings} and \ref{sec:branch}. As in Figure \ref{fig:basins}, the logarithmic magnitude of the dispersion relation is plotted on the complex domain of the search variable. Nine roots and five isolated singularities are visible, while the branch point exists at $\beta^2 = 6.25$ with a branch cut that was selected to extend towards $-i\infty$. The five contours are shown in brown, numbered from right to left, since the fundamental mode is the rightmost root. Each contour encloses one unique singularity, which restricts the number of enclosed roots to a maximum of three in this example. The contours overlap to ensure sufficient coverage of the search domain, while the branch point is avoided by shrinking the adjacent contours appropriately. The branch cut is repositioned to minimize conflict with the contours and the Riemann sheet which contains the bound and interesting leaky modes is chosen.}
\label{fig:rootcircles}
\end{center}
\end{figure}

Having obtained the locations of the singularities, we may select circular contours that satisfy the requirements of Section \ref{sec:selection}. The entire selection process is summarized in the form of pseudocode in Algorithm \ref{alg:contour}, and an example is displayed in Figure \ref{fig:rootcircles}. We roughly center each contour on each singularity, with boundaries that extend three-quarters of the distance to adjacent singularities. Such contours exhibit modest overlap, which is necessary for circular contours to provide adequate coverage of the complex plane. This arrangement also ensures the boundaries do not stray too close to neighboring singularities, which would negatively affect the numerical convergence of the integral. More explicitly, the center $c$ and radius $r$ of the $k$th contour is defined by
\begin{align}
c_k &= \frac{1}{4} u_k + \frac{3}{8} u_{k-1} + \frac{3}{8} u_{k+1}, & r_k &= \frac{3}{8} (u_{k+1} - u_{k-1}),
\label{eq:contours}
\end{align}
where $u_k$ is location of the $k$th unique singularity, numbered from $k=1$ onwards. For example, we may assign $u_1$ as $\alpha_c a = 0$, while all other $u_k$ are given by \eqref{eq:sings} beginning with $j_{m,1}$. No singularity can be assigned to $u_0$, but $u_0$ serves only to determine the lower bound of the first contour. Only one root is ordinarily located in this region, as typified by Figure \ref{fig:rootcircles}, and we find that a definition such as $u_0 = 2u_1 - u_2$ is satisfactory for locating this root.

\begin{algorithm}[t]
\SetAlgoLined
\SetKwFunction{ArgumentPrincipleMethod}{ArgumentPrincipleMethod}
\SetKw{Call}{call}
\KwIn{parameters of dispersion relation \eqref{eq:disprel}}
\KwOut{roots of dispersion relation \eqref{eq:disprel}}
determine singularities using \eqref{eq:sings}\;
\lIf{$m \neq 0$}{append singularity at $\alpha_c a = 0$}
define contours according to \eqref{eq:contours}\;
\For{each contour}{
\lIf{contour includes branch point \eqref{eq:branch}}{shrink contour}
determine enclosed singularities and known roots\;
\Call \ArgumentPrincipleMethod\;
\If{a root is close to singularity}{
define a smaller contour enclosing the singularity and root\;
\Call \ArgumentPrincipleMethod\;
}
polish roots using an iterative method;
}
collect, sort, and return roots\;
\caption{Defining contours and driving the argument principle method}
\label{alg:contour}
\end{algorithm}

A further detail is that roots of \eqref{eq:disprel} may be found very close to singularities, for example $(root - singularity)/singularity < 10^{-2}$. Since we use iterative methods to polish the roots, and such methods diverge from singularities, the argument principle method must locate these roots to relatively high accuracy to ensure successful subsequent polishing. The necessary accuracy may be achieved by a finer discretization of the contours \eqref{eq:contours}, but in our experience a more efficient alternative is to define a much smaller contour tightly encircling the singularity and its nearby root. Such a contour can accurately predict the root location despite being coarsely discretized.

\subsection{Branch points and branch cuts}
\label{sec:branch}
Critical to the success of the method is identifying the branch points that exist within the search domain, which must be avoided by the argument principle method. Furthermore, branch cuts also emanate from the branch points, which must be relocated to avoid interfering with the contours \eqref{eq:contours}. Note however that branch points do not always appear, so we also briefly discuss conditions for their appearance.

The square root in the definition of $\alpha_b$, \eqref{eq:alpha}, has the potential to introduce a branch point. As discussed in Section \ref{sec:selection}, the sign of $\alpha_c a$ is inconsequential, but the sign of $\alpha_b a$ is not, since it enters into $H_m(z)$, which is neither an odd nor even function. Furthermore, $H_m(z)$ has its own branch point at $z=0$.\autocite{abramowitz1964handbook}, thus creating a logarithmic branch point at
\begin{equation}
(\alpha_b a)^2 = (k^2\epsilon_b\mu_b - \beta^2)a^2 = 0.
\label{eq:branch}
\end{equation}
However, this branch point only exists if $\alpha_b a$ varies, that is, if either $k$, $\beta$, $\epsilon_b$, $\mu_b$, or $a$ is the search variable. If $\alpha_b a$ is invariant, which is the case if either $\epsilon_c$ or $\mu_c$ is the search variable, then no branch points are present, and the techniques described in this section are unnecessary.

As with the contour selection process, we again use physical insight into our dispersion relation \eqref{eq:disprel} to simplify the handling of the branch point. Only a single branch point exists within the search domain, assuming for example $\beta^2$ is used as the search variable rather than $\beta$. The branch point corresponds to the light line of the background medium, and it is known that solutions are located near the branch point for only a brief parameter range near cutoff. The exception is the fundamental mode, which has no cutoff and instead asymptotes towards the light line.\autocite{snyder1983optical,okamoto2006fundamentals} Only this mode is usually found in the vicinity of the branch point for a wide parameter range, but fortunately reliable asymptotic formulas exist for its location.\autocite{chang2007strong} Hence, the vast majority of the search domain is unaffected by the branch point, and the few modes near the branch point can be found by alternative means. We thus adopt the simple method of adjusting the contours to avoid the branch point rather than more sophisticated methods such as conformal mapping, which unnecessarily complicate the implementation.\autocite{smith1991numerical,smith1992mode,smith1993unfolding,bakhtazad1997general} 

When the contours \eqref{eq:contours} are chosen, only one or possible two of these will enclose the the branch point. We shrink the sizes of these particular contours to exclude the branch point. Nevertheless, it is still desirable that these contours pass as close as possible to the branch point, to ensure that they reliably capture roots in this region. Since the branch point is also a singularity, this introduces an unavoidable point of slow convergence during numerical quadrature. Switching from the equidistant trapezoidal method to adaptive Simpson's method greatly ameliorates this issue, to be described in Section \ref{sec:discrete}.


Next, the branch cuts that radiate from the branch point must be repositioned. Two branch cuts exist, which separately originate from either the square root and Hankel functions. In MATLAB, and most other numerical packages, these are positioned along the negative real axis in the domains of each function, which usually corresponds precisely to the region containing the bound fiber modes. Secondly, the relevant branch of the square root function must be chosen to locate the desired roots, whether they be the bound, leaky, or radiative modes.


We move the square root branch cut by defining a new function, $\sqroot(z,\phi)$, given by
\begin{equation}
\sqroot(z, \phi) = \pm\sqrt{z e^{-i\phi}} e^{i\phi/2},
\label{eq:sqrt}
\end{equation}
where $\phi$ defines the desired anti-clockwise rotation of the branch cut in the complex plane and $\sqrt{z}$ is the original numerical implementation of the square root function. The appropriate sign to choose in \eqref{eq:sqrt} is informed by sign of $\imag(\alpha_b)$ necessary to ensure either decay or growth at infinity in the background medium. For $H_m(z)$, this is the upper and lower half plane, respectively. The branch cut of $H^{(1)}_m(z)$ can be moved by exploiting the identities between various Bessel functions,\autocite{abramowitz1964handbook}
\begin{align}
H^{(2)}_m(z) &= (-1)^{m+1} H^{(1)}_m(-z), & K_m(z) &= \frac{\pi}{2} i^{m+1} H^{(1)}_m(iz),
\label{eq:hankelbranch}
\end{align}
where $H^{(2)}_m(z)$ is the Hankel function of the second kind, and $K_m(z)$ is the modified Bessel of the second kind. Due to the rotations of the argument of $H^{(1)}_m(z)$ in \eqref{eq:hankelbranch}, these identities effectively rotate the branch cut of $H^{(1)}_m(z)$ by $\pi$ and $\pi/2$ respectively. We choose to use the second of \eqref{eq:hankelbranch}, which conveniently includes the domains of both bound and the interesting leaky modes.


In addition to the singularities already treated in Section \ref{sec:selection}, singularities may also be present in the immediate vicinity of the branch point. Note however, that these singularities are also absent if the branch point is absent. This last set of singularities arises from the complex zeros of $H_m(z)$, which on the principal branch lies just below the negative real axis and in an eye-shaped region below the origin.\autocite{abramowitz1964handbook} The former are eliminated during the movement of the branch cut, so can be neglected. The few that remain exist at relative small values of $\alpha_b a$, so are tightly clustered around the branch point, unlike the singularities of $J(\alpha_c a)$ which extend across the complex plane. To deflate the contributions of the corresponding singularities, asymptotic formulas exist for the approximate locations of the zeros of $H_m(z)$ using the Olver expansion,\autocite{abramowitz1964handbook,doring1966complex} or alternatively the argument principle method can be applied to $H_m(z)$. However, the region of the complex plane containing these singularities is less physically interesting, as no bound modes are located here while the leaky modes exhibit explosive exponential growth. Thus, we choose to neglect this region.


\subsection{Discretization and evaluation}
\label{sec:discrete}
Using the contours selected in Section \ref{sec:sings} and modified in Section \ref{sec:branch} to avoid branch cuts, the argument principle method is ready to be applied. Either the moments $s_k$, \eqref{eq:moments}, or their derivative free equivalents $t_k$, \eqref{eq:derivfree}, are evaluated, which proceeds via discretization and numerical quadrature. We detail the procedure for $s_k$, which is presented as pseudocode in Algorithm \ref{alg:argprin}, while the procedure for $t_k$ follows with minor adjustments summarized below.

The equipartition trapezoidal rule is applied to the circular contours \eqref{eq:contours}, which are parameterized as
\begin{equation}
z_C(t) = c + re^{2\pi it},
\label{eq:circle}
\end{equation}
where $c$ and $r$ are the center and radius of each contour, and $t \in [0,1]$. The interval is covered by $M$ equidistant points defined by $t_j = j/M$. To improve numerical conditioning, the polynomial weight function $z_k$ in \eqref{eq:moments} and \eqref{eq:derivfree} is replaced by a shifted and scaled coordinate system
\begin{equation}
w =  \frac{z-c}{r},
\label{eq:scaleshift}
\end{equation}
which has magnitude $1$ along the boundary of the contour \eqref{eq:circle}. This has the additional advantage that the powers of $w$ are simple, $w^n = e^{2n\pi it}$. 

Applying the trapezoidal rule to \eqref{eq:moments}, a discretized version is obtained,
\begin{equation}
s_n = \frac{r}{M} \sum_{j=1}^M e^{2\pi i(n+1)t_j} \frac{f'(z(t_j))}{f(z(t_j))}.
\label{eq:nums}
\end{equation}
The moments are deflated of singularities and known roots using the scaled and shifted version of \eqref{eq:deflate}, obtained using \eqref{eq:scaleshift}. The locations of roots within the contour are retrieved using Newton's identities \eqref{eq:newid}. These locations are then used as initial guesses for Newton's method \eqref{eq:newton}, which typically converges to double precision within 1 or 2 iterations. 

Numerical implementation of the derivative free version \eqref{eq:derivfree} is similar. An expression almost identical to \eqref{eq:nums} is obtained, but without the $f'(z(t_j))$ factor. Since deflation of singularities is unnecessary and deflation of known roots is more involved than \eqref{eq:deflate}, we neglect this step. The roots are retrieved from the moments using the generalized eigenvalue equation \eqref{eq:geneig}. Root polishing is performed with a derivative free method, for which we choose the secant method.

\begin{algorithm}[t]
\SetAlgoLined
\SetKw{Or}{or}
\KwIn{contour center and radius, known roots and singularities}
\KwOut{roots within contour}
\Repeat{$s_0$ is equal to integer within tolerance \Or number of points exceeds limit}{
initialize or double the number of quadrature points\;
evaluate $s_0$ using trapezoidal rule, \eqref{eq:nums}\;
}
\If{$s_0$ has not yet converged}{
continue numerical quadrature using adaptive Simpson's method\;
}
determine number of enclosed roots from $s_0$\;
evaluate remaining moments, $s_n$, using trapezoidal or Simpson's\;
deflate singularities and known roots using \eqref{eq:deflate}\;
construct polynomial using Newton's identities, \eqref{eq:newid}\;
find roots of polynomial using standard routine\;
\caption{The argument principle method}
\label{alg:argprin}
\end{algorithm}

Since the subsequent polishing via iterative methods is much more computationally efficient than the argument principle method, the moments \eqref{eq:nums} should only be accurate enough to ensure the rapid convergence of the chosen iterative method. Obtaining further accuracy using finer discretization of the contour is unnecessary and inefficient. However, the accuracy of the roots obtained from a particular contour cannot be predicted at the time of discretization, since the presence of roots close to the contour drastically slows the convergence of the trapezoidal rule. Thus, the number of points $M$ necessary cannot be predefined, and needs to be varied for each contour. To maintain the efficiency of the numerical implementation, we adaptively increase $M$, beginning with $M=32$, until the desired accuracy is estimated to be reached. A good estimate for this purpose is the numerical value of $s_0$, which is known to be an integer.\autocite{delves1967numerical} When $s_0$ differs from an integer by less than a specified tolerance, no further discretization is necessary and the algorithm proceeds. Alternatively, a good criterion for derivative free moments \eqref{eq:derivfree} is the convergence of $t_0$.


Finally, the trapezoidal rule becomes inappropriate to use if a root or singularity lies very close to a contour, as convergence is then unacceptably slow. For example, as few as 32 quadrature points may be necessary to achieve single precision accuracy, or as many as $2^{15}$. Since the root locations are usually unknown when the contours are set, we begin with the trapezoidal rule, and switch to the adaptive Simpson's rule if more than 512 points are necessary. Preexisting function evaluations are reused, and convergence of $s_0$ or $t_0$ to the desired accuracy is usually obtained with only a few additional function evaluations. Numerical testing shows that even for contours lying within $10^{-7}$ of a root or singularity, no more than 1024 points are necessary. To maximize efficiency, we implement adaptive Simpson's rule iteratively rather than recursively, which is the traditional implementation. This preserves the function evaluations of $f(z)$ and $f'(z)$, which are reused during the evaluation of all other moments.

\section{Results and discussion}
\label{sec:numerics}
\begin{figure}[!t]
\begin{center}
\subfloat{\includegraphics{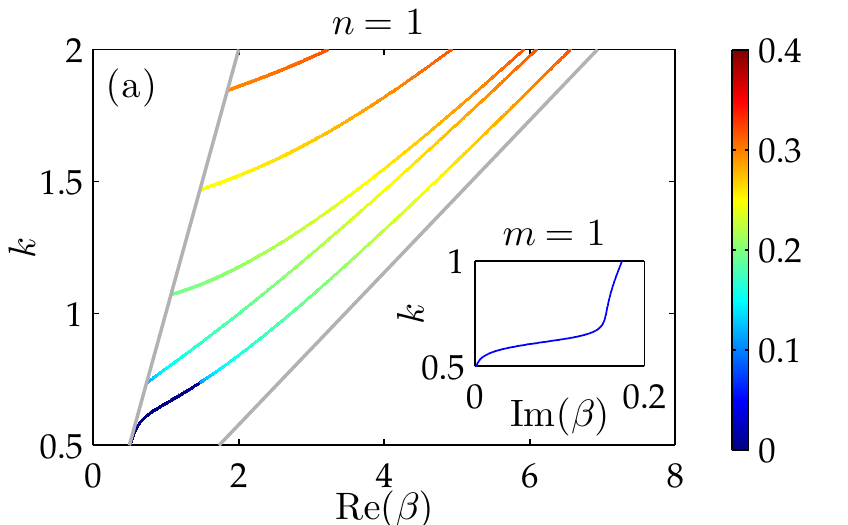}}\\
\noindent\makebox[\textwidth]{%
\subfloat{\includegraphics{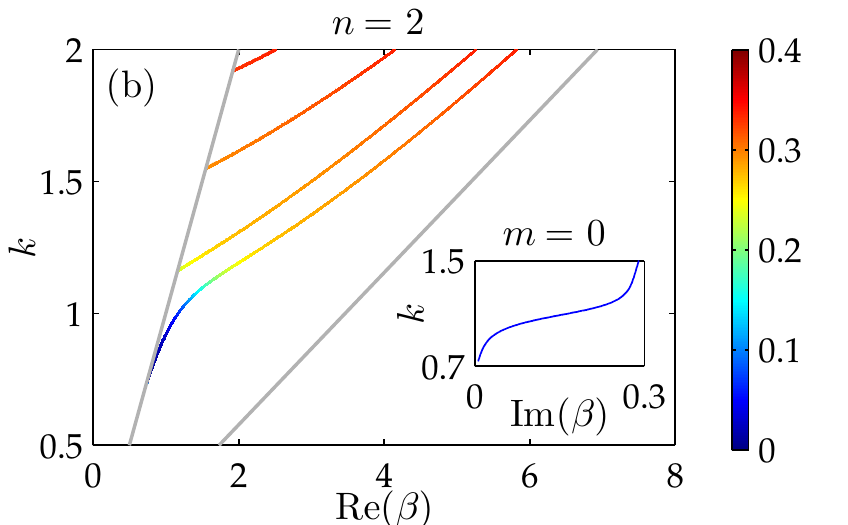}}
\subfloat{\includegraphics{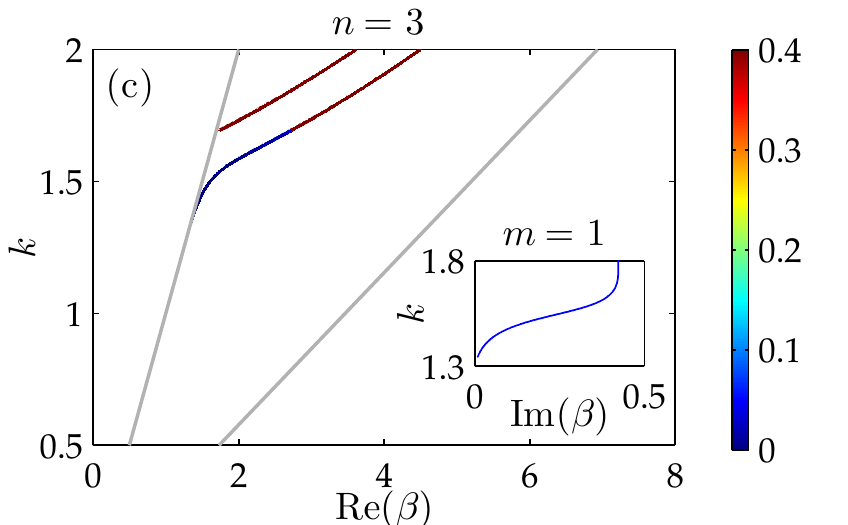}}
}
\caption[]{Dispersion relations produced by the method using complex $\beta$ as the search variable, with $\imag(\beta)$ represented by color. Cylinder permittivity is $\epsilon_c = 12+i$ and the background is vacuum, while cylinder radius $a$ normalized to $1$. The gray line on the left of each figure indicates the light line of vacuum, while the other gray line corresponds to $k=\sqrt{12}\beta$. Each subfigure plots modes with the same radial order $n$, indicated by the figure titles. Within each figure, the modes are arranged from smallest angular order to largest. The exceptions are the two lowest angular orders within each figure, with $m=1$ being the fundamental mode and $m=0$ being the second mode for (a) and (c). All modes shown are ``bound'' in the sense that fields that decay exponentially to zero in the background, though energy does radiate from the fiber due to the complex $\beta$. Only modes where $\real(\beta)$ falls below the light line are shown. The insets of each figure show the rapid increase of $\imag(\beta)$ of the fundamental modes that asymptote towards the light line, with the inset titles indicating the angular order of this mode.}
\label{fig:dielectric}
\end{center}
\end{figure}

We now demonstrate the numerical implementation on \eqref{eq:disprel} under two modes of operation, with either $\beta$ or $\epsilon_c$ as the search variable. The two cases have different numerical and physical characteristics, the primary distinction being that $\beta$ case features a branch point but not the $\epsilon_c$ case. In both cases, all eigenmodes can be identified by two quantum numbers: $m$ the azimuthal order, and $n$ the radial order, which count the number of nodes in the respective directions. Different orders $m$ are solutions to different dispersion relations \eqref{eq:disprel}, while different orders $n$ are different solutions to the same dispersion relation. For radial orders $n$, the mode with the fewest nodes is the lowest order mode, which we number from $1$ to $\infty$. Meanwhile, orders $m$ extend from $-\infty$ to $\infty$. However, the symmetry of the cylinder, and consequently the dispersion relation \eqref{eq:disprel}, ensures that negative $m$ modes have the same eigenvalue while the field pattern is merely inverted, $m \rightarrow -m$. We thus need to consider only orders $m \geqslant 0$.

We begin with the $\beta$ eigenvalue search, focusing on the intermediate frequency region, where no satisfactory approximate solutions to the dispersion relation are available. For example, derivations of asymptotic expressions for the long wavelength limit are available,\autocite{chang2007strong} which may be generalized for our purposes
\begin{equation}
\beta^2 - k^2\epsilon_b\mu_b \approx \frac{2}{a^2} \exp\left(\frac{\epsilon_b\mu_b + \epsilon_c\mu_c + \epsilon_b\mu_c + \epsilon_c\mu_b}{\epsilon_b\mu_b(\epsilon_c\mu_c-\epsilon_b\mu_b)(ka)^2}\right).
\end{equation}
Meanwhile, the geometric optics limit, representing the opposite limit with large $ka$, has been extensively treated in textbooks.\autocite{snyder1983optical,okamoto2006fundamentals}

We perform the root search for a lossy high index contrast fiber, with $\epsilon_c = 12 + i$, where all eigenvalues become complex. We operate on $\beta^2$ as the search variable, which creates only a single branch point and generates a simpler arrangement of roots and singularities for the placement of contours, as seen is Figure \ref{fig:rootcircles}. For plotting purposes, we switch to $\beta$, presenting several dispersion relations in Figure \ref{fig:dielectric} of various radial and angular orders. Note that wavenumber $k$ is the independent variable and $\beta$ is the dependent search variable, but real part of $\beta$ is plotted on the horizontal axis to to conform to the literature convention. The imaginary part of $\beta$ is indicated by color. 

The light line is a key feature, given by $\beta^2 = k^2\epsilon_b\mu_b$, shown as the gray line on the left. For lossless fibers, this separates the bound modes with real $\beta$ from the leaky modes with complex $\beta$. For complex $\beta$ eigenmodes, the light line remains an important feature, as the lowest mode of each radial order asymptotes towards the light line. These modes have low propagation losses when close to the light line, but $\imag(\beta)$ rapidly increases away from the light line. This increase is displayed in the insets of each figure. Note that the identity of this fundamental mode does not remain constant across different radial orders, alternating between the first and zeroth angular orders. For all the other complex $\beta$ modes, the light line no longer represents the cutoff, whereby $\alpha_b$ changes from real to imaginary values. The modes no longer intersect the light line in the complex plane, and instead $\alpha_b$ remains complex while transitioning smoothly as a function of $k$ into modes with large propagation losses with large $\imag(\beta)$. Thus, we neglect to show these segments of the dispersion relations as such modes are less physically interesting. If desired though, these modes are easily located by our method.

\begin{figure}[!t]
\begin{center}
\subfloat{\includegraphics{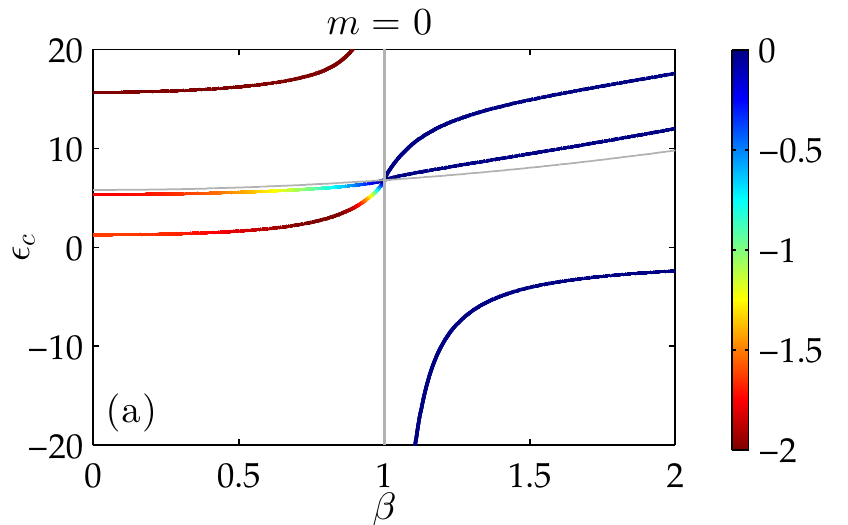}}\\
\noindent\makebox[\textwidth]{%
\subfloat{\includegraphics{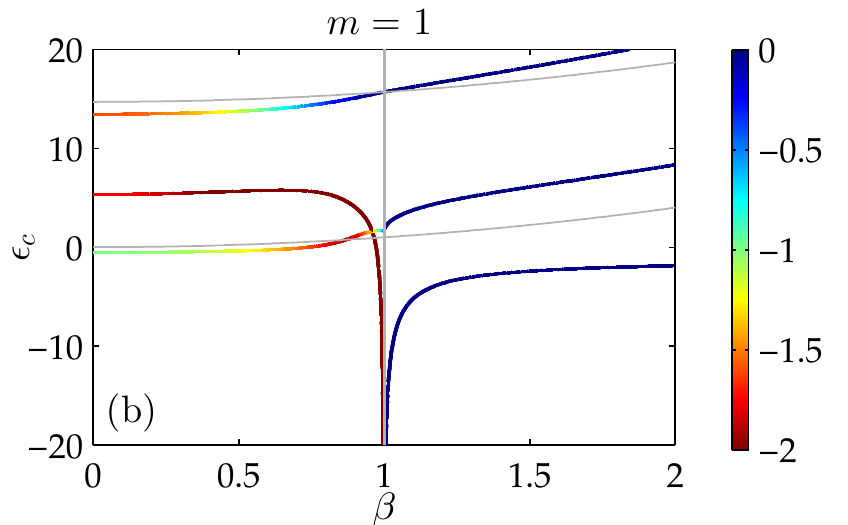}}
\subfloat{\includegraphics{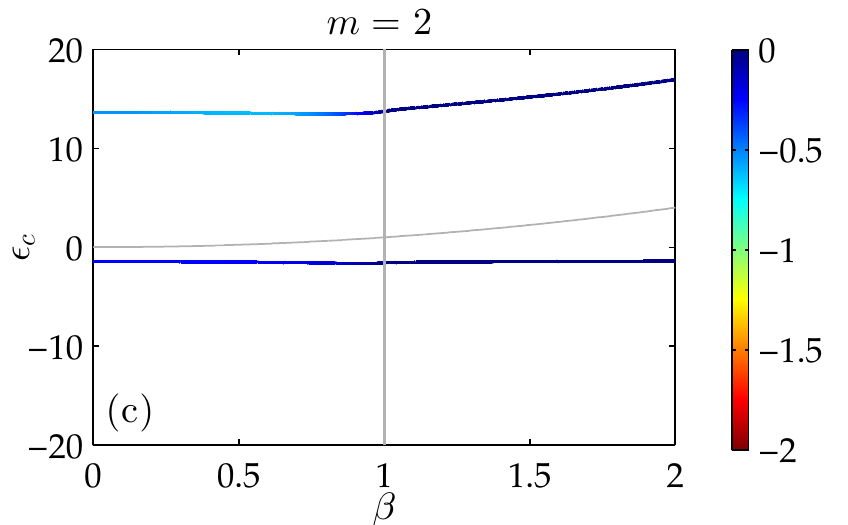}}
}
\caption[]{Dispersion relations with complex cylinder permittivity $\epsilon_c$ as the search variable, with $\imag(\epsilon_c)$ indicated by color. Wavenumber is $k=1$ and cylinder radius $a$ is normalized to 1. Unlike Figure \ref{fig:dielectric}, the subfigures plot modes with a common angular order, given by the title, of differing radial orders. The vertical gray line represents the light line, while the curved gray lines are singularities of the dispersion relation \eqref{eq:disprel}. Bound modes with real $\epsilon_c$ exist to the right of the light line, while complex $\epsilon_c$ radiative modes exist to the left. Unlike Figure \ref{fig:dielectric}, where only physically relevant modes on a specific Riemann sheet are shown, the complete set of modes within the given parameter range are displayed.}
\label{fig:eps}
\end{center}
\end{figure}

Next, we consider the case of $\epsilon_c$ as the eigenvalue. While such modes are less commonly sought than the preceding case, there is emerging need from methods that expand the inhomogeneous Maxwell equations in terms of eigenmodes. Here, the frequency and the background index are set, $\beta$ is the independent variable of the dispersion relation, and $\epsilon_c$ is the dependent search variable. We plot in Figure \ref{fig:eps} these dispersion relations, this time uniting modes of the same angular order in each subfigure. Note that the light line is a vertical line since frequency is fixed, and is not a source of branch points. Bound modes exist to the right of the light line and always have real eigenvalues, while radiative modes exist to the left and always have complex eigenvalues. The latter satisfy the Sommerfeld boundary condition and decay at infinity. To ensure that the steady state is maintained despite radiating energy, $\epsilon_c$ has gain to compensate. Thus, the physical significance of these modes is less important than their mathematical properties, which is to provide a complete basis for expanding any source or excitation.

All data points within the dispersion relations were produced using our principle argument method and polished using Newton's method. The roots were not traced, in other words solutions at one frequency were not used as initial guesses for the next frequency. This demonstrates the robustness of the method across the parameter range, with no risk of losing roots due to roots interacting with each other or with singularities. Despite the seemingly mundane nature of such interactions, it has been our experience that they pose many difficulties for the design of robust root search methods that rely only on iterative methods. For example, Newton's method can cope with some of these issues with clever choice of initial guesses, but not under all circumstances. The argument principle method alleviates all such difficulties, and effortlessly delivers the complete set of complex modes. Indeed, we prefer to use the method even when all roots are known to be real, because of its simplicity and robustness.

The two step procedure is efficient, as most contours only require coarse discretization to ensure the roots are successfully and rapidly polished using Newton's method. Typically, only 64 or 128 evaluation points are necessary for the argument principle method to find the roots to more than 5 digits of accuracy, despite each contour covering a significant area surrounding each singularity, \eqref{eq:contours}. This remains true even for roots that lie close to singularities, where greater accuracy is necessary to ensure successful polishing. Highly accurate predictions can be generated by employing a smaller but still coarsely discretized contour. The exceptions are contours that pass near the singular branch point, which require many function evaluations near the branch point. However, less than 1000 points are required, even for contours that lie within $10^{-7}$ of the branch point. This number is still far from problematic for our inexpensive to evaluate dispersion relation, \eqref{eq:disprel}. Nevertheless, if better performance is desired, asymptotic formulas can be used instead for the few roots that asymptote towards the branch point. Alternatively, a logarithmic mapping of the complex plane can move the singularity to infinity,\autocite{bakhtazad1997general} which also allows a more thorough search of the region around the branch point.


All dispersion relations were generated using the moments \eqref{eq:moments}, rather than the derivative free moments \eqref{eq:derivfree}. Our numerical testing revealed that the derivative free moments, in conjunction with the generalized eigenvalue problem \eqref{eq:geneig}, was more accurate for real roots, but was less accurate for complex roots. This was particularly true for roots with narrow attraction basins due to their larger larger $f'(z)$ and therefore smaller weights in \eqref{eq:derivfree}. Since these are the very roots that require more accurate initiation to polish, we preferred to use the moments \eqref{eq:moments} throughout.

\section{Conclusion}
\label{sec:conclusion}
We have successfully implemented the argument principle method, specialized for the step index fiber dispersion relation. We demonstrate its robust global convergence, reliably locating all desired roots, whether real or complex, within a specified region of the complex plane. The method is insensitive to issues which plague iterative root search methods, arising from difficult attraction basins caused by roots interacting with singularities, branch points, and with other roots. The method is fast and efficient, requiring few function evaluations to produce accurate estimates of roots that are subsequently rapidly polished by Newton's method.

In contrast to previous implementations, we allow search domains that include isolated singularities. The locations and orders of all such singularities are first found analytically, and their contributions to the moments of the argument principle method may be deflated. Furthermore, the presence of branch points is not fatal to the method. We avoid all non-meromorphic points while still capturing all desired roots by employing a contour selection procedure informed by a little knowledge of our dispersion relation. To achieve this, a few contours that pass near the branch point are necessary, and the resulting poor convergence during numerical quadrature due to the singularity is managed by switching to adaptive quadrature. Since we select the contours once and for all at the outset, we obtain a simple implementation which avoids all the complexities of other implementations during the process of repartitioning and resizing the of search domain. Moreover, we avoid the hazardous interaction of this process with branch points.

We operate our implementation in two ways, finding eigenmodes with either complex propagation constant or cylinder permittivity as the eigenvalue. These meet two different needs of the current literature, for modeling propagation along high index contrast or lossy fibers, and for expanding sources using source-free eigenmodes. All solutions of the dispersion relation can be generated, including bound, lossy, and radiative modes, both real and complex. Our argument principle implementation is a useful tool particularly for the challenging but important intermediate frequency region between the long wavelength and geometric optics limits.

\section*{Acknowledgments}
PYC acknowledges fellowship support from the Tel Aviv University Center for Nanoscience and Nanotechnology.

\appendix
\section{Retrieving roots from moments}
Two methods are available to obtain the locations of the roots from the moments. Newton's identities apply to the moments \eqref{eq:moments} and their deflated equivalents, while the generalized eigenvalue equation also applies to the derivative free moments \eqref{eq:derivfree}.

\subsection{Newton's identities}
\label{sec:newid}
A polynomial of order $N$ with the same roots as the target transcendental equation, \eqref{eq:transcend}, can be obtained 
\begin{equation}
\prod_{k=1}^N (z-z_k) = \sum_{j=0}^N (-1)^{N+j} e_{N-j} z^j,
\label{eq:newpoly}
\end{equation}
where $e_i$ are the elementary symmetric polynomials
\begin{equation}
\begin{aligned}
e_0 &= 1, & e_1 &= z_1 + z_2 + \ldots + z_N,\\
e_2 &= \sum_{1 \leqslant i < j \leqslant N} z_i z_j, 
&\quad\quad \hdots \quad\quad
e_N &=  z_1 z_2 \ldots z_N.
\end{aligned}
\end{equation}
Newton's identities relate these polynomials to the moments \eqref{eq:deflate} 
by a recursive formula\autocite{ralston2001first}
\begin{equation}
ke_k = \sum_{n=1}^k (-1)^{n-1} e_{k-n} s_n^*,
\label{eq:newid}
\end{equation}
which begins with $e_1 = s_1^*$, thus allowing the polynomial \eqref{eq:newpoly} to be constructed and solved.

\subsection{Generalized eigenvalue equation}
\label{sec:geneig}
The locations of the roots can be retrieved, even when weighted by the unknown $f'(z)$ from \eqref{eq:derivfree}, by solving the generalized eigenvalue problem
\begin{equation}
H^<x = \lambda Hx,
\label{eq:geneig}
\end{equation}
where eigenvalues $\lambda$ are the locations of the roots, the two Hankel matrices are
\begin{align}
H &= \begin{bmatrix}
t_0 & \cdots & t_{R-1} \\
\vdots & \ddots & \vdots \\
t_{R-1} & \cdots & t_{2R-2}
\end{bmatrix}, &
H^< &= \begin{bmatrix}
t_1 & \cdots & t_R \\
\vdots & \ddots & \vdots \\
t_R & \cdots & t_{2R-1}
\end{bmatrix},
\end{align}
and $R$ is the number of enclosed roots. Note that twice the number of moments need to be evaluated compared to the procedure using Newton's identities, but this is usually an insignificant numerical burden, since evaluation of $f(z)$ is typically consumes the most time. Indeed, time savings are usually realized through use of the derivative free formulation if the evaluation of $f'(z)$ is numerically expensive.

\section{Derivatives}
The argument principle method and subsequent Newton's method are most efficient when explicit derivatives are available. We evaluate these explicitly for both $\beta^2$ and $\epsilon_c$ as the search eigenvalue. To simplify the algebraic manipulations, we introduce the symbols
\begin{align}
R^J_m &= \frac{J'_m(\alpha_c a)}{\alpha_c a J_m(\alpha_c a)}, & R^H_m &= \frac{H'_m(\alpha_b a)}{\alpha_b a H_m(\alpha_b a)}.
\end{align}
The dispersion relation can be rewritten as 
\begin{equation}
(\mu_c R^J_m - \mu_b R^H_m) (\epsilon_c R^J_m - \epsilon_b R^H_m) - \left(\frac{m\beta}{k}\right)^2 \left(\frac{1}{(\alpha_c a)^2}-\frac{1}{(\alpha_b a)^2}\right)^2 = 0.
\label{eq:disprelr}
\end{equation}

The bulk of the calculation involves the derivative
\begin{equation}
\begin{aligned}
\frac{\partial}{\partial \alpha_c} R^J_m &= \frac{1}{(\alpha_c a J_m(\alpha_c a))^2} [\alpha_c a^2 J_m(\alpha_c a) J''_m(\alpha_c a) - J'_m(\alpha_c a)(\alpha_c a^2 J'_m(\alpha_c a) + a J_m(\alpha_c a))]\\
&= -\frac{1}{\alpha_c J_m(\alpha_c a)^2} \left[J_m(\alpha_c a)^2 + \frac{2}{\alpha_c a} J_m(\alpha_c a) J'_m(\alpha_c a) - J_{m+1}(\alpha_c a) J_{m-1}(\alpha_c a)\right].
\end{aligned}
\label{eq:drj}
\end{equation}
The defining Bessel differential equation was used in the second equality. Since the Hankel function obeys the same identities, the derivative $\partial R^H_m/\partial \alpha_b$ is obtained by the substitution $\alpha_c \rightarrow \alpha_b$ and $J_m(\alpha_c a) \rightarrow H_m(\alpha_b a)$ in \eqref{eq:drj}.

First consider the derivative of \eqref{eq:disprelr} with respect to $\epsilon_c$. In order use \eqref{eq:drj}, the chain rule is applied to derive
\begin{equation}
\begin{gathered}
\mu_c\frac{\partial R^J_m}{\partial \alpha_c} \frac{\partial \alpha_c}{\partial \epsilon_c} [\epsilon_c R^J_m - \epsilon_b R^H_m] + [\mu_c R^J_m - \mu_b R^H_m] \left[R^J_m + \epsilon_c \frac{\partial R^J_m}{\partial \alpha_c} \frac{\partial \alpha_c}{\partial \epsilon_c}\right]\\
- \frac{4m^2\beta^2}{\alpha^3k^2a^2} \frac{\partial \alpha_c}{\partial \epsilon_c} \left(\frac{1}{(\alpha_c a)^2} - \frac{1}{(\alpha_b a)^2}\right),
\end{gathered}
\end{equation}
where
\begin{equation}
\frac{\partial \alpha_c}{\partial \epsilon_c} = \frac{k^2 \mu_c }{2 \alpha_c}.
\end{equation}

Now consider the derivative with respect to $\beta^2$, giving
\begin{equation}
\begin{gathered}
\left[\mu_c\frac{\partial R^J_m}{\partial \alpha_c} \frac{\partial \alpha_c}{\partial \beta^2} - \mu_b\frac{\partial R^H_m}{\partial \alpha_b} \frac{\partial \alpha_b}{\partial \beta^2}\right] [\epsilon_c R^J_m - \epsilon_b R^H_m] 
+ [\mu_c R^J_m - \mu_b R^H_m] \left[\epsilon_c \frac{\partial R^J_m}{\partial \alpha_c} \frac{\partial \alpha_c}{\partial \beta^2} - \epsilon_b \frac{\partial R^H_m}{\partial \alpha_b} \frac{\partial \alpha_b}{\partial \beta^2}\right] \\
- \left(\frac{m}{k}\right)^2\left(\frac{1}{(\alpha_c a)^2} - \frac{1}{(\alpha_b a)^2}\right)^2\left[1+2(\beta a)^2 \left(\frac{1}{(\alpha_c a)^2} + \frac{1}{(\alpha_b a)^2} \right)\right],
\end{gathered}
\end{equation}
where
\begin{equation}
\frac{\partial \alpha}{\partial \beta^2} = - \frac{1}{2\alpha}.
\end{equation}

\printbibliography
\end{document}